\documentclass[aps,pra,showpacs,showkeys,reprint,twocolumn,superscriptaddress,preprintnumbers,amsmath,amssymb]{revtex4-1}
\usepackage{graphicx}
\usepackage{dcolumn}
\usepackage{bm}
\usepackage{xspace} 
\usepackage{dsfont} 
\usepackage{psfrag}
\usepackage{bbm}
\usepackage{hyperref}
\usepackage{float}
\usepackage[caption=false]{subfig}
\usepackage[usenames]{color}
\usepackage{textcomp}

\begin{document}

\title{Bright Integrated Photon-Pair Source for Practical Passive Decoy State Quantum Key Distribution}
\author{S. Krapick}
\email[corresponding author: ]{krapick@mail.uni-paderborn.de}
\author{M. S. Stefszky}
\affiliation{Department of Physics, University of Paderborn, Warburger Str. 100, 33098 Paderborn, Germany}
\author{M. Jachura}
\affiliation{Department of Physics, University of Paderborn, Warburger Str. 100, 33098 Paderborn, Germany}
\affiliation{Faculty of Physics, University of Warsaw, Ho$\dot{z}$a 69, 0zero-681 Warsaw, Poland}
\author{B. Brecht}
\affiliation{Department of Physics, University of Paderborn, Warburger Str. 100, 33098 Paderborn, Germany}
\author{M. Avenhaus}
\affiliation{Department of Physics, University of Paderborn, Warburger Str. 100, 33098 Paderborn, Germany}
\affiliation{Max-Planck-Institute for the Science of Light, G\"unther-Scharowsky-Str. 1, 91058 Erlangen, Germany}
\author{C. Silberhorn}
\affiliation{Department of Physics, University of Paderborn, Warburger Str. 100, 33098 Paderborn, Germany}
\affiliation{Max-Planck-Institute for the Science of Light, G\"unther-Scharowsky-Str. 1, 91058 Erlangen, Germany}
\date{\today}

\begin{abstract}
We report on a bright, nondegenerate type-I parametric down-conversion source, which is well suited for passive decoy-state quantum key distribution. We show the photon-number-resolved analysis over a broad range of pump powers and we prove heralded higher-order $n$-photon states up to $n=4$. The inferred photon click statistics exhibit excellent agreements to the theoretical predictions. From our measurement results we conclude that our source meets the requirements to avert photon-number-splitting attacks.
\end{abstract}


\maketitle

Quantum key distribution (QKD) has been a field of high interest for almost three decades now, since it allows two trustworthy parties, Alice and Bob, to communicate with unconditional security under certain constraints. However, realistic implementations of QKD schemes suffer from security loopholes \cite{Lydersen2010,Fung2007,Makarov2005,Qi2007,Lamas-Linares2007,Gottesman2004,Tang2013a} due to technical imperfections. Among these loopholes, the photon-number-splitting (PNS) attack \cite{Brassard2000} allows an eavesdropper, Eve, to take advantage of non-ideal properties of real-world photon sources. In particular, photon-pair sources based on parametric down-conversion (PDC) emit signals with higher-order photon contributions, which could be intercepted and stored by Eve to gain information about the secret key during the classical phase of a standard prepare-and-measure protocol \cite{Bennett1984}. When Eve replaces a lossy quantum channel with a perfect one, she could mimic the detected click statistics of the lossy quantum channel and cannot be detected.

In $2007$ Mauerer $et\, al.$ proposed the passive decoy-state QKD protocol \cite{Mauerer2007} and theoretically showed the circumvention of photon number splitting attacks, even in the presence of imperfect photon-pair sources based on PDC. It was also proven that the unconditionally secure transmission distance is on par with perfect single-photon sources \cite{Mauerer2007,Curty2010}. The scheme is based on the idea of intermittent decoy states \cite{Hwang2003,Lo2005} within the quantum key string in order to detect Eve's presence. But, in contrast to active decoy schemes, the passive decoy approach turns the unavoidable higher photon-number components in a PDC into a real benefit by tagging them as intrinsic decoys. Importantly, this does not require the active modulation or phase randomization of the photon source's emission. The passive decoy scheme offers the distinct advantage that the PDC-based system itself does not open any side channels with distinguishing information for different intensities, because all required decoy states are post-selected \textit{after} transmission. However, the implementation of a reliable, bright, compact and efficient PDC source with a well-known photon-number distribution is crucial to achieve the desired performance.

In this paper we present a robust and bright integrated photon-pair source based on titanium-indiffused, periodically poled waveguide structures in lithium niobate (Ti:PPLN). It efficiently generates signal photons at around $803$ nm and idler photons around $1573$ nm. The former lend themselves to efficient photon-number-resolved detection, whereas the latter allow for low-loss transmission in fiber-based QKD systems. Our source is capable of splitting the generated pairs on chip in a spatio-spectral manner \cite{Krapick2013}. This conveniently allows Alice to keep one half (signal) of the strictly correlated pairs for thorough photon-number analysis, whereas the other half (idler) can be transmitted to her trusted counterpart Bob.

The key feature of our source is that it meets the requirements for practical passive decoy-state QKD. In particular, we measured its click statistics using photon-number-resolving detectors and inferred the photon statistics from the measurement. We registered higher-order PDC photon numbers reliably and demonstrate heralded $n$-photon states up to $n=4$, which can be employed as post-selectable decoys in order to prevent PNS attacks. In the following, we verify that the detection statistics of our source behaves as expected for the different $n$-photon states. This will prove its applicability as a basic building block in highly secure QKD systems based on passive decoy-state selection.

First, we write down the probability that Bob's binary detector generates a click from an arbitrary $m$-photon state:
\begin{equation}
\label{eq:01}
p(\mathrm{click})=1-(1-\eta_\mathrm{B})^m,
\end{equation}
with Bob's overall transmission and detection efficiency $\eta_\mathrm{B}=\eta_\mathrm{C}\cdot\eta_\mathrm{OC}\cdot\eta_\mathrm{Det}$. Herein $\eta_\mathrm{C}$ denotes the length-dependent quantum channel efficiency, $\eta_\mathrm{OC}$ is the transmission of supplementary optical components and $\eta_\mathrm{Det}$ labels Bob's detector efficiency. Note that Eq. (\ref{eq:01}) implies different click probabilities for different $m$-photon states.

Second, we assume that Alice's photon-number-resolving detector yields $n$ photon detection events from an $m$-photon state with $m\ge n$. Consequently, the conditioned probability for a click at Bob's detector is

\begin{equation}
\begin{split}
\label{eq:02}
& p(\mathrm{click}|n)=\frac{p(\mathrm{click}\cap n)}{p(n)}\\
& =\frac{\sum\limits_{m=n}^\infty\dbinom{m}{n}\rho_{m}\eta_\mathrm{T}^{n}\left(1-\eta_\mathrm{T}\right)^{m-n}\left(1-\left(1-\eta_\mathrm{B}\right)^{m}\right)}{\sum\limits_{m=n}^\infty\dbinom{m}{n}\rho_{m}\eta_\mathrm{T}^{n}\left(1-\eta_\mathrm{T}\right)^{m-n}},
\end{split}
\end{equation}
where $p(\mathrm{click}\cap n)$ is the cumulative joint probability of a click event in Bob's detector from an $m$-photon state, while $n$ out of $m$ photons impinge on Alice's photon-number-resolving detector. The coefficients $\rho_\mathrm{m}$ describe the photon-number distribution of the $m$-photon state, which is Poissonian \cite{Mauerer2009,Helwig2009} for the spectrally multi-mode PDC sources like those expected in our case. The term $\eta_\mathrm{T}$ labels the overall efficiency of the photon-number-resolving detector and includes losses as well as the genuine detection efficiency. The efficiency $\eta_\mathrm{T}$ for our case is experimentally accessible as will be described below in Eq. (\ref{eq:06}).

Especially, in the limit of a low efficiency, $\eta_\mathrm{B}\ll1$, we can calculate the conditioned probabilities $p(\mathrm{click}|n)$ for different $n$-photon states from Eq. (\ref{eq:02}), and find that the approximation
\begin{equation}
\label{eq:03}
r\left(n\right)=\frac{p\left(\mathrm{click}|n\right)}{p\left(\mathrm{click|1}\right)}\approx\frac{n\cdot\eta_\mathrm{B}}{\eta_\mathrm{B}}=n,
\end{equation}
is valid. This means that the fractions of Bob's click probabilities conditioned on different $n$ and the click probability conditioned on the $one$-photon contribution scale approximately linear with $n$. At higher efficiencies $\eta_\mathrm{B}$ the values of $r\left(n\right)$ will decrease. Thus, at elevated pump power levels the impact of higher-order photon contributions as well as non-ideal photon-number-resolving detector properties must be taken into account.
\begin{figure}[h]
\includegraphics[width=\linewidth]{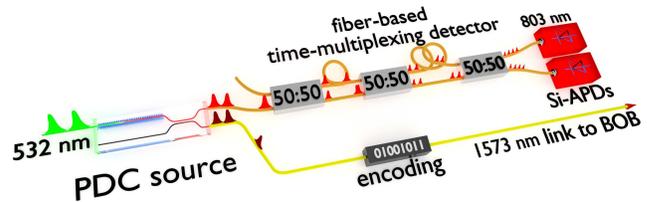}
\caption{\label{fig:01} Alice's source configuration for passive decoy-state QKD: PDC signal photons ($803$ nm) are separated on chip from idler photons ($1573$ nm), fed into the eight-bin time-multiplexer, and are detected with binary detectors (Si-APDs). The (encoded) idler photons are transmitted to Bob via the quantum channel.}
\end{figure}
In order to realize photon-number resolution with Alice's detection apparatus, we implemented a time-multiplexing detector \cite{Achilles2003,Fitch2003} (TMD) for signal wavelengths of $803$ nm. The delay between individual time bins ($\sim127$ ns) is set larger than the dead time of standard silicon avalanche photodiodes used for signal photon detection. The chosen architecture provides eight temporal output modes behind the TMD (photon numbers of $n\le8$) without losing photons by dead-time effects. Due to this limitation, we additionally take convolution effects \cite{Achilles2004,Achilles2006a} of higher-order photon contributions into account for the data analysis method \cite{Avenhaus2008} as well as for the theoretical predictions. A schematic of one possible passive decoy-state QKD implementation at Alice's side including the TMD is shown in Fig. \ref{fig:01}.

For the experimental analysis, we derive the conditioned probabilities $p\left(\mathrm{click}|n\right)$ as the fraction of measured click events in Bob's detector, given that an $n$-photon state is detected by Alice's non-ideal TMD, and the total number of events:
\begin{equation}
\label{eq:04}
p\left(\mathrm{click}|n\right)=\frac{N(\mathrm{click}|n)}{N(\mathrm{click}|n)+N(\mathrm{no\,click}|n)}.
\end{equation}
Note, that $p\left(\mathrm{click}|n\right)$ still has to be corrected for convolution effects.

\begin{figure}
\includegraphics[width=\linewidth]{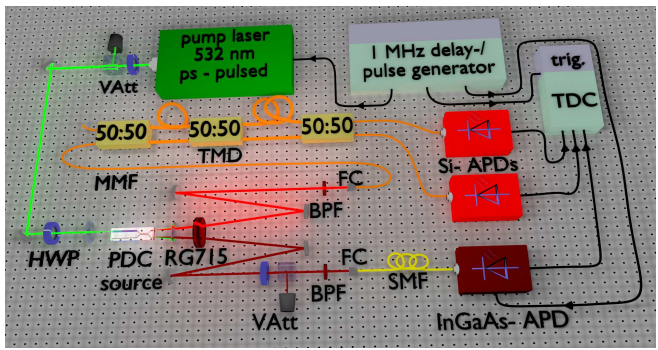}
\caption{\label{fig:02} Experimental implementation of photon-number-resolving PDC analysis (VAtt: variable attenuator; HWP: half-wave plate; FC: fiber coupling; RG715: home-coated absorber; BPF: band pass filter; SMF: single-mode fiber; MMF: multi-mode fiber; TMD: time-multiplexing detector; APD: avalanche photodiode; TDC: time-to-digital-converter); see text for details.}
\end{figure}

We carry out pump-power-dependent measurements with the setup shown in Fig. \ref{fig:02}, since the mean photon number of  PDC states is related to the power of the pump pulse. Our pump laser offers ps pulses at $532$ nm, which can be variably attenuated and coupled into our periodically poled waveguide structure. Generated signal and idler photons are demultiplexed on chip and separated into two output beams. Behind the waveguide chip, we clean up PDC photons from background and residual pump light using a home-coated absorber and narrowband dielectric filters. In the signal arm we address the TMD and, subsequently, two free-running silicon avalanche photodiodes (APD), both with $\eta_\mathrm{Si}=0.55$ detection efficiency. The idler arm consists of a variable attenuator, which mimics an arbitrary channel loss $\eta_\mathrm{C}$. Behind this device, we address a gated InGaAs-APD, which offers $2.5$ ns detection windows, a detection efficiency of $\eta_{\mathrm{Det}}=0.24$ and $1\,\mu$s dead time. It exhibits a dark count probability of $p_{\mathrm{dc}}=1.75\times10^{-4}$ per gate.

For the detection, all APDs are connected to a time-to-digital converter (TDC) offering $82$ ps resolution. A home-programmed software analyzes the impinging signals for coincidences in order to extract the PDC click statistics at a specific pump power, i.e., mean photon number. Pump laser, TMD, and InGaAs-APD are synchronized, triggered, and delay compensated in terms of optical path differences by a delay generator running at $1$ MHz repetition rate.

We analyze the photon-number-resolved click statistics of our PDC process at pump powers that range over two orders of magnitude. A high pump power forces the generation of higher-order photon states. The maximum accessible cw-equivalent pump power of $2\,\mu$W corresponds to a mean photon number of $\langle n \rangle=0.84$, and it is determined by the saturation limit of our data acquisition system. At each pump-power level we can set arbitrary quantum channel transmission $0<\eta_{\mathrm{C}}\le 1$ in the idler arm, mimicking different transmission distances of a real-world QKD system. 

The Klyshko efficiencies \cite{Klyshko1980} of our signal and idler arm $\eta_\mathrm{T}$ and $\eta_\mathrm{B}$ are given by the ratio of coincidence counts $N_\mathrm{coinc}$ and the total number of single counts in the respective opposite arm, $N_\mathrm{B}$ and $\sum_{n\ge1}N_\mathrm{T}\left( n \right)$. In order to correct the Klyshko efficiencies for uncorrelated events, we estimate the number of accidentals beforehands as
\begin{equation}
\label{eq:05}
N_\mathrm{acc}=\frac{N_\mathrm{B}\cdot\sum_{n\ge1} N_\mathrm{T}(n)}{N_\mathrm{Trig}},
\end{equation}
where $\sum_{n\ge1} N_\mathrm{T}(n)$ is the accumulated number of detection events in the TMD, $N_\mathrm{B}$ denotes the number of click events in the InGaAs-APD and $N_\mathrm{Trig}$ is the number of trigger events within our measurement time. This correction provides us with a lower bound for the actual Klyshko efficiency, because higher-order photon contributions lead to an overestimation of the real values. We also subtract coincidences $N_\mathrm{dc,B}$ and $N_\mathrm{dc,T}$ caused by dark counts of the respective detector, and we finally find
\begin{equation}
\label{eq:06}
\eta_\mathrm{T}=\frac{N_\mathrm{coinc}-N_\mathrm{acc}-N_{\mathrm{dc,T}}}{N_\mathrm{B}}
\end{equation}
for the Klyshko efficiency in the signal/TMD arm and
\begin{equation}
\label{eq:07}
\eta_\mathrm{B}=\frac{N_\mathrm{coinc}-N_\mathrm{acc}-N_{\mathrm{dc,B}}}{\sum_{n\ge1} N_\mathrm{T}\left( n\right)}
\end{equation}
for the Klyshko efficiency of the idler arm.

The results for different pump powers are shown in Table \ref{tab:01} and indicate that our calculated Klyshko efficiencies are only reliable at low pump powers, since we tend to over-estimate accidentals - according to Eq. (\ref{eq:05}) - for increasing pump powers and, thus, will artificially decrease the Klysko efficiencies. In order to predict the behavior of different $n$-photon states in the following, we base our theoretical calculations on Klyshko efficiencies obtained at the lowest available pump power. Note that this will surely underestimate the influence of accidentals.
\begin{table}[h]
\caption{\label{tab:01} Overview of the power-dependent Klyshko efficiencies at maximum channel transmission}
\begin{ruledtabular}
\begin{tabular}{rrrrrrrr}
$P_\mathrm{p}$ [nW]&20&50&100&200&500&1000&2000\\\hline
$\eta_\mathrm{B}[\%]$&10.75&10.72&10.65&10.02&9.97&9.41&8.55\\
$\eta_\mathrm{T}[\%]$&17.76&17.76&17.72&17.66&17.21&16.58&15.47
\end{tabular}
\end{ruledtabular}
\end{table}

In order to ensure distinct detection probabilities for different $n$-photon states after transmission through the quantum channel, we analyze the click statistics and reconstruct the probabilities $p\left(\mathrm{click}|n\right)$ therefrom using Eq. (\ref{eq:04}) and the inverse convolution matrix \cite{Achilles2004,Coldenstrodt-Ronge2007,Avenhaus2008} of our TMD. A typical measurement result at the highest accessible pump power of $2\;\mu$W and with $\eta_\mathrm{C}=1$ is shown in Table \ref{tab:02}. We can clearly identify click events up to photon numbers $n=4$ within $60$ s of measurement time, which strongly indicates heralded four-photon states. 
\begingroup
\begin{table*}
\caption{\label{tab:02} Typical click-statistics at $2\;\mu$W pump power and with channel transmission $\eta_\mathrm{C}=1$.}
\begin{ruledtabular}
\begin{tabular}{rrrrrr}
&zero-photon&one-photon&two-photon&three-photon&four-photon\\ \hline
$N(\mathrm{'no\,click'}|n)$&49244089&6157356&334960&10383&197\\
$N(\mathrm{click}|n)$&3049176&1092105&102653&4608&112\\
$N_\mathrm{T}(n)$&52293265&7249461&437613&14991&309
\end{tabular}
\end{ruledtabular}
\end{table*}
\endgroup

\begin{figure}
\includegraphics[width=0.95\linewidth]{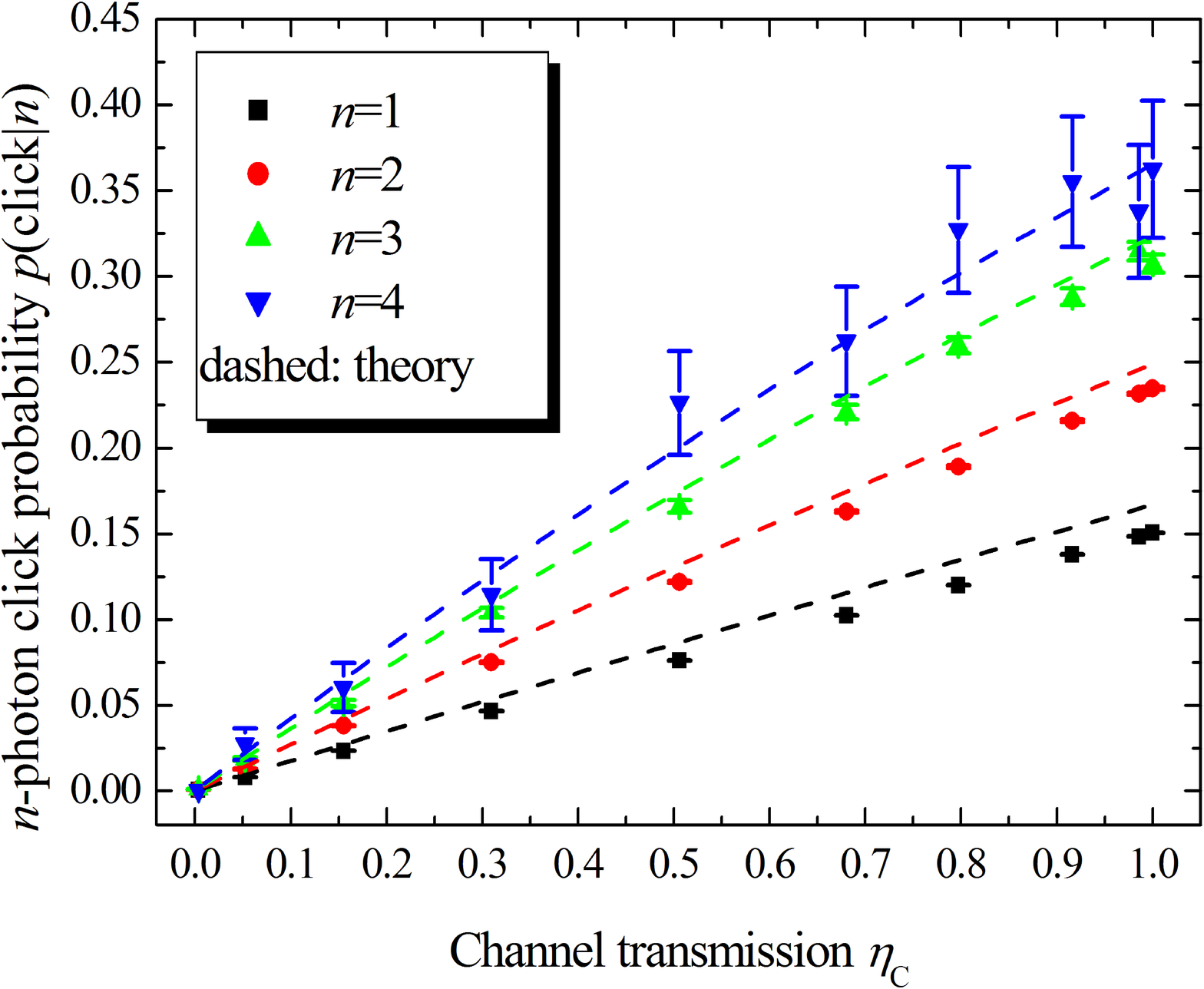}
\caption{\label{fig:03} Dependencies of the $n$-photon click probabilities on the channel transmission at $2\;\mu$W cw-equivalent pump power; dashed: theory curves.}
\end{figure}

In Fig. \ref{fig:03} we plot $p\left(\mathrm{click}|n\right)$ versus the channel transmission $\eta_\mathrm{C}$, the latter of which represents an arbitrarily long transmission device between the two QKD parties. The distinct $n$-photon components clearly follow different slopes and show also higher detection probabilities for higher photon states. This verifies not only the high brightness of our source, but it also agrees excellently with the expected behavior. Our measurements closely match the theoretical curves calculated with Eq. (\ref{eq:02}), where we considered Poissonian distributions $\rho_\mathrm{m}$ as well as convolution effects. We also assumed Klyshko efficiencies of the low power regime, $\eta_\mathrm{T}=0.1776$ and $\eta_\mathrm{B}=0.1075$, respectively. Note, that the differences to the detection efficiencies $\eta_\mathrm{Si}$   and $\eta_\mathrm{Det}$ are caused by losses introduced by the implemented optical components.

In a practical passive decoy-state QKD system, a PNS attack will be detected by Alice and Bob, since the click probability of one-photon contributions would be increased artificially compared to the above statistics. Thus, even if Eve was able to replace parts of the lossy quantum channel by a perfect one for the attack, her presence can be recognized. This is due to the fact, that it is still undecided during transmission, which subset of $n$-photon number states will be employed as decoys. With our analysis scheme it is easy for Alice to anticipate, how the click statistics at Bob's side should behave at distinct pump powers. Thus, our PDC source fulfills the necessary requirements for passive decoy-state QKD in terms of predictable photon statistics and accessibility to higher-order $n$-photon states in general. The seemingly growing mismatch between experimental data and theory curves for small $n$ at large $\eta_\mathrm{C}$ can be explained by uncorrellated coincidences, which have an impact on $p\left(\mathrm{click}|n\right)$. As stated above, by applying only Klyshko efficiencies from the low-power measurement to the theory, we underestimate accidentals for higher pump powers.

\begin{figure}[h]
\includegraphics[width=0.95\linewidth]{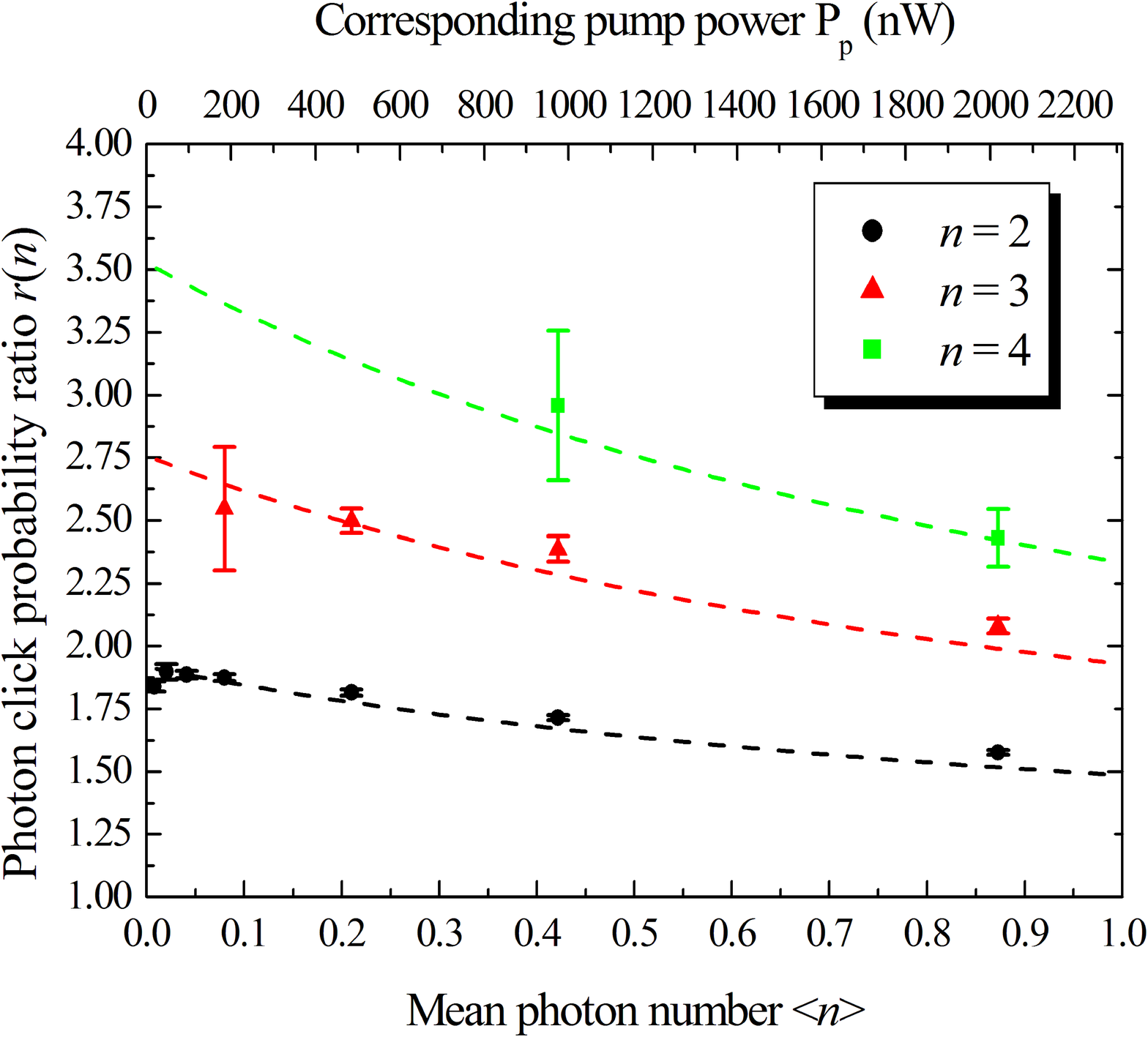}
\caption{\label{fig:04} The click probability ratios $r\left(n\right)$ follow Poissonian distributions (dashed: theoretical predictions for $\eta_\mathrm{T}=0.1776$ and $\eta_\mathrm{B}=0.1075$).}
\end{figure}

The applicability of our source over a large range of different brightnesses, as needed for optimization of the passive decoy scheme, is shown by the behavior for different mean photon numbers of the PDC states. In particular, we calculated $r\left( n \right)$ according to Eq. (\ref{eq:03}) from the individual deconvoluted click probabilities at variable channel transmissions. The average ratios are plotted against the mean photon number $\langle n \rangle$ in Fig. \ref{fig:04}. We did not register significant three- and four-photon components at small mean photon numbers within acceptable measurement durations. However, our measurement data exhibit decreasing $r\left( n \right)$ at higher pump powers according to Poissonian statistics. This, on one hand, underlines the photon-number-resolving capabilities of our TMD, while on the other hand the very good agreement to the theory proves that the $n$-photon PDC states in our spectrally broad source (FWHM$\left(803\,\mathrm{ nm}\right)\sim0.7$ nm) show almost pure Poissonian distributions. Remaining deviations from theory can be explained by the finite number of spectral modes in our source, and again by the impact of uncorrelated accidentals at higher mean photon numbers. The distinct detection probability ratios for different $n$-photon states are key to accessing higher-order photons as decoy states reliably. A PNS attack will change the above characteristics in a way that the ratios $r\left(n\right)$ decrease artificially due to the increased amount of one-photon contributions during classical PNS attacks and, thus, can be detected. Therefore, Fig. \ref{fig:04} shows that our source is usable at a broad range of pump powers.

In summary, we have shown the suitability of our integrated photon-pair source for practical passive decoy-state QKD. We determined its photon-number-resolved emission characteristics. Our results prove distinct detection probabilities for different $n$-photon states up to $n=4$ at arbitrary quantum channel efficiencies. The dependencies of the $n$-photon state click probability ratios on increasing mean photon numbers closely follow Poissonian distributions according to the spectral multi-mode character of our PDC. Together with the capability to reliably provide heralded four-photon states and the excellent agreement to theoretical predictions, the high brightness of our source fulfills the requirements for passive decoy-state QKD. This constitutes an important step towards real-world implementation of QKD schemes, where all security loopholes have to be eliminated.

The authors thank the Deutsche Forschungsgemeinschaft for funding this work within the Graduate Program on 'Micro-and Nanostructures in Optoelectronics and Photonics' (GRK 1464 II).

\bibliographystyle{apsrev4-1}
\bibliography{Literatur-Database}

\end{document}